# Statistical tests for a sequence of random numbers by using the distribution function of random distance in three dimensions

Sergii Koliada
February 21, 2014

**Abstract:** The distribution function of a random distance in three dimensions is given and some new three-dimensional $d^2$-tests of randomness are suggested. We show that our test statistics are not correlated with the usual test statistics and are therefore an other promising way to determine the quality of generated random numbers.

## 1. Introduction

For the test of uniform randomness of a given sequence of numbers, several test procedures are known [2], [4]. Frequency tests, random walk tests and tests based on Matusita's distance are used for the test of equiprobability [3]. Run tests, gap tests, combination tests, serial correlation tests and $d^2$-tests[1] are used for the test of independence [1].

In this paper, we suggest new three-dimensional $d^2$-tests that may be considered to be an extension of usual two-dimensional $d^2$-tests. We will denote the two-dimensional $d^2$-tests and our three-dimensional $d^2$-tests by $D_2^2$-tests and $D_3^2$-tests respectively.

In Section 2 we calculate the distribution function of the three-dimensional random distance. In Section 3 we propose $D_3^2$-tests. In Section 4 we calculate the correlation coefficients between a $D_3^2$-test statistic and each of the other test statistics by the Monte Carlo methods and compare $D_3^2$-tests with the other tests. In Section 5 we discuss some variations and claim the superiority of $D_3^2$-tests to $D_2^2$-tests for testing the three-dimensional structure of random numbers.

## 2. The distribution of the three-dimensional random distance

Suppose $(X_i, \ldots, X_k)$ is a k-dimensional random variable whose elements are mutually independent and have the uniform distribution on [0, 1]. Let it be called a k-dimensional random vector.

Now we define the k-dimensional random distance or the Euclid distance $D_k$ between two random vectors $(X_i, \ldots, X_k)$ and $(X_i', \ldots, X_k')$ by

$$(2.1) \qquad D_k = \sqrt{\sum_{i=1}^{k} (X_i - X_i')^2}.$$

Let $F_k(a^2)$ be the distribution function of the square of the k-dimensional random distance $D_k^2$. $F_k(a^2)$ has the reduction formula given by

$$\Pr\{D_k^2 \leqq a^2\} = F_k(a^2) = \begin{cases} 0 & (a^2 \leqq 0) \\ F_k^i(a^2) & (i-1 \leqq a^2 \leqq i), (i = 1, 2, \ldots, k) \\ 1 & (k \leqq a^2) \end{cases}$$

(2.2)

---
[1] Additional examples for this test can be found at: arxiv.org/abs/1304.4786

where

$$\begin{cases} F_k^i(a^2) = \int_0^{a^2-i+1} F_{k-1}^i(a^2-t)(t^{-1/2}-1)\,dt \\ \qquad\qquad + \int_{a^2-i+1}^1 F_{k-1}^{i-1}(a^2-t)(t^{-1/2}-1)\,dt,\, (i=1,2,\ldots,k) \\ F_{k-1}^0(a^2) = 0 \\ F_{k-1}^k(a^2) = 1 \end{cases} \quad (2.3)$$

and

$$(2.4) \qquad F_1^1(a^2) = 2a - a^2 \qquad (0 \leqq a^2 \leqq 1).$$

The mean and variance of $D_k^2$ are

$$(2.5) \qquad E(D_k^2) = \frac{k}{6}, \qquad V(D_k^2) = \frac{7k}{180}.$$

The distribution of $D_2^2$ is known to have the form (for example, see [4])

$$F_2(a^2) = \begin{cases} \pi a^2 - \frac{8}{3}a^3 + \frac{1}{2}a^4 & (0 \leqq a^2 \leqq 1) \\ \frac{1}{3} + (\pi - 2)a^2 + 4(a^2-1)^{1/2} \\ \quad + \frac{8}{3}(a^2-1)^{3/2} - \frac{1}{2}a^4 - 4a^2\sec^{-1}a & (1 \leqq a^2 \leqq 2). \end{cases}$$

(2.6)

The distribution function of $D_3^2$ is given by

$$F_3(a^2) = \begin{cases} \frac{4}{3}\pi a^3 - \frac{3}{2}\pi a^4 + \frac{8}{5}a^5 - \frac{1}{6}a^6 & (0 \leqq a^2 \leqq 1) \\ \left(\frac{5}{2}\pi + \frac{43}{30}\right) - 6(a^2-1)^{1/2} \\ \quad + \left(3\pi + \frac{7}{2}\right)(a^2-1) - \frac{8}{3}\pi a^3 \\ \quad - 10(a^2-1)^{3/2} + \frac{5}{2}(a^2-1)^2 \\ \quad - \frac{16}{5}(a^2-1)^{5/2} + \frac{1}{3}(a^2-1)^3 \\ \quad + 6a^4\sec^{-1}a & (1 \leqq a^2 \leqq 2) \\ \left(\frac{23}{2}\pi - \frac{343}{30}\right) + 14(a^2-2)^{1/2} \\ \quad + \left(9\pi - \frac{21}{2}\right)(a^2-2) + 10(a^2-2)^{3/2} \\ \quad + \left(\frac{3\pi-5}{2}\right)(a^2-2)^2 + \frac{8}{5}(a^2-2)^{5/2} \\ \quad - \frac{1}{6}(a^2-2)^3 \\ \quad - 2(3a^4 + 6a^2 - 1)\sec^{-1}\sqrt{a^2-1} \\ \quad + 8a^3\sec^{-1}(a^2-1) - \frac{8}{3}\pi a^3 & (2 \leqq a^2 \leqq 3). \end{cases}$$

(2.7)

The derivation of (2.7) can be easily found at [4][2]. Table 1 gives numerical values of

---

[2] Similar online at: www.math.kth.se/~johanph/habc.pdf („The probability distribution of the distance between two random points in a box"), page 10, (22)

$F_3(a^2)$ for $a^2=0(0.1)3$.

| $a^2$ | $F_3(a^2)$ | $a^2$ | $F_3(a^2)$ |
|---|---|---|---|
| 0.1 | 0.0902303 | 1,6 | 0.9947872 |
| 0.2 | 0.2134496 | 1,7 | 0.9971309 |
| 0.3 | 0.3385455 | 1,8 | 0.9984757 |
| 0.4 | 0.4569491 | 1,9 | 0.9992138 |
| 0.5 | 0.5648731 | 2 | 0.9996066 |
| 0.6 | 0.6604814 | 2,1 | 0.9998121 |
| 0.7 | 0.7429195 | 2,2 | 0.9999160 |
| 0.8 | 0.8118854 | 2,3 | 0.9999656 |
| 0.9 | 0.8674103 | 2,4 | 0.9999875 |
| 1 | 0.9097346 | 2,5 | 0.9999962 |
| 1,1 | 0.9398577 | 2,6 | 0.9999991 |
| 1,2 | 0.9607864 | 2,7 | 0.9999997 |
| 1,3 | 0.9750998 | 2,8 | 0.9999999 |
| 1,4 | 0.9846709 | 2,9 | 1.0000000 |
| 1,5 | 0.9908888 | 3 | 1.0000000 |

Table 1: Numerical values of $F_3(a^2)$ for $a^2=0(0.1)3$.

When $k \geq 4$, the distribution function $F_k(a^2)$ can not be expressed by elementary functions. However, if k is large, $D_k^2$ is asymptotically normally distributed and we can expand the distribution function of

(2.8) $$Z_k = \frac{D_k^2 - \frac{k}{6}}{\sqrt{\frac{7k}{180}}},$$

up to the terms of order $k^{-3/2}$:

(2.9) $$\Pr\{Z_k \leqq z\} = F_k^*(z) = \varphi(z) - \frac{1}{\sqrt{2\pi}} e^{-z^2/2} \left\{ \frac{\beta_3}{6\sqrt{k}} h_2(z) \right.$$
$$+ \frac{\beta_4}{24k} h_3(z) + \frac{\beta_3^2}{72k} h_5(z)$$
$$+ \frac{\beta_5}{120k^{3/2}} h_4(z) + \frac{\beta_3 \beta_4}{144k^{3/2}} h_6(z)$$
$$\left. + \frac{\beta_3^3}{1296k^{3/2}} h_8(z) \right\} + o\left(\frac{1}{k^{3/2}}\right)$$

with

(2.10) $$\varphi(z) = \int_{-\infty}^{z} \frac{1}{\sqrt{2\pi}} \exp\left(-\frac{t^2}{2}\right) dt,$$

(2.11)
$$\begin{cases} \beta_3 = \frac{88}{49}\sqrt{\frac{5}{7}} \\ \beta_4 = \frac{606}{343} \\ \beta_5 = -\frac{8160}{3773}\sqrt{\frac{5}{7}} \end{cases}$$

and

(2.12)
$$\begin{cases} h_2(z) = z^2 - 1 \\ h_3(z) = z^3 - 3z \\ h_4(z) = z^4 - 6z^2 + 3 \\ h_5(z) = z^5 - 10z^3 + 15z \\ h_6(z) = z^6 - 15z^4 + 45z^2 - 15 \\ h_8(z) = z^8 - 28z^6 + 210z^4 - 420z^2 + 105. \end{cases}$$

## 3. $D_3^2$-tests

The results of the previous section make it possible to construct some test statistics for testing the null hypothesis that the numbers $x_1, \ldots, x_{6n}$ are the observed values of a random sample of size $6n$ from the uniform distribution on $[0, 1]$.

To this end, we divide the whole sequence into a sequence of $2n$ triplets; $(x_1, x_2, x_3)$, $\ldots$ , $(x_{6n-2}, x_{6n-1}, x_{6n})$. Considering them as $2n$ observed values of random vectors in the unit cube, we can calculate $n$ squares of the Euclid distance between the vectors $(x_{6m-5}, x_{6m-4}, x_{6m-3})$ and $(x_{6m-2}, x_{6m-1}, x_{6m})$ given by

(3.1) $\quad d_3^2 = (x_{6m-5} - x_{6m-2})^2 + (x_{6m-4} - x_{6m-1})^2 + (x_{6m-3} - x_{6m})^2$, $(m = 1, 2, \ldots, n)$.

Further, we divide the interval $0 \leq a^2 \leq 3$ into $p$ subintervals $(a_{i-1}^2, a_i^2]$, $(i=1, 2, \ldots, p)$, where a's are rather arbitrary and count the number $f_i$ of components of $d_3^2$ falling into the i-th interval. Since the expected number $F_i$ of components of $d_3^2$ falling into this interval is equal to

(3.2) $\quad F_i = n[F_3(a_i^2) - F_3(a_{i-1}^2)]$, $(i=1, 2, \ldots, p)$,

the statistic

(3.3)
$$Q_3^2 = \sum_{i=1}^{p} \frac{(f_i - F_i)^2}{F_i}$$

gives a $D_3^2$-test.

In fact, under the null hypothesis, the statistic $Q_3^2$ is approximately distributed according to the chi-square[3] distribution with $p-1$ degrees of freedom, and we can reject the hypothesis when $Q_3^2$ is too large. In this paper, we set p=10 and selected the subintervals in a way that $F_i$ is nearly equal to 0.1n for any i (i=1, 2, … , p).

---

[3] Additional informations to this and similar tests can be found at:
academic.csuohio.edu/zhao_w/teaching/Old/EEC685-F05/lecture18-web-2up.pdf,
yatani.jp/HCIstats/, koliada.co.nf/stat-papers/ or http://www.nyx.net/~tmacfarl/STAT_TUT/stat_tut.ssi

| i | $(a_{i-1}^2, a_i^2]$ | $F_3(a_i^2)-F_3(a_{i-1}^2)$ |
|---|---|---|
| 1 | (0.000, 0.100] | 0.0902303 |
| 2 | (0.100, 0.182] | 0.1005155 |
| 3 | (0.182, 0.261] | 0.0994948 |
| 4 | (0.261, 0.343] | 0.1003229 |
| 5 | (0.343, 0.430] | 0.0999724 |
| 6 | (0.430, 0.526] | 0.1004274 |
| 7 | (0.526, 0.634] | 0.0990487 |
| 8 | (0.634, 0.767] | 0.1006073 |
| 9 | (0.767, 0.950] | 0.0995822 |
| 10 | (0.950, 3.000] | 0.1097985 |

Table 2: Numerical values for $(a_{i-1}^2, a_i^2]$ and $F_3(a_i^2)-F_3(a_{i-1}^2)$

## 4. Comparison among $D_3^2$-tests and usual tests by the Monte Carlo methods

In this section, we compare $D_3^2$-tests with usual tests. For this purpose we calculated the correlation coefficients between a $D_3^2$-test statistic and each of the seven usual test statistics by the Monte Carlo methods.

In these experiments, the interval (0, 1] is divided into 10 subintervals of equal length to construct the test statistic based on Matusita's distance and the frequency test statistic, runs are divided into three to construct the run test statistic, the gap test statistic and the combination test statistic are made so as to have three degrees of freedom, the serial correlation test statistic is made to be a t-test statistic concerned with the correlation coefficients between a random number and the next one, and the $D_2^2$-test statistic is to be an analogue of the $D_3^2$-test statistic.

We used three arithmetic random numbers (a), (b) and (c) that had been accepted by the usual seven tests:

(4.1)
(a) $R_n = 134459 R_{n-1}$ (mod $2^{22}$), $R_0=28551$
(b) $R_n = 65539 R_{n-1}$ (mod $2^{31}$), $R_0=153249823$
(c) $R_n = 243 R_{n-1}$ (mod $10^8$), $R_0=50249347$

Using 993 random numbers, we calculated the simulated value of each test statistics, and repeated it 200 times and thus got 200 simulated values to each test statistics. We calculated the correlation coefficients between the $D_3^2$-test statistic, $Q_3^2$, and each of the seven test statistics. The results are shown in Table 3.

| Test statistic \\ Random numbers | (a) | (b) | (c) |
|---|---|---|---|
| Frequency | 0.0847 | -0.0517 | 0.0011 |
| Run | -0.0156 | -0.0947 | -0.0598 |
| Combination | 0.0604 | 0.0695 | 0.0623 |
| Gap | -0.0627 | -0.0528 | 0.0192 |
| Serial correlation | -0.0086 | -0.1064 | 0.0283 |
| Matusita's distance | 0.0778 | -0.0516 | -0.0028 |
| $Q_{2(12)(34)}^2$ | 0.0198 | 0.0404 | -0.0130 |

Table 3: The correlation coefficients between $Q_3^2$ and each of the other seven test statistics

From the results, we may say that $D_3^2$-test statistics are approximately not correlated with the other test statistics. Since $D_3^2$-tests and $D_2^2$-tests are alike in the form, we examined their relation carefully. And we had the following remarkable results.

When we calculate $d_2^2$ from the four consecutive random numbers $x_{4m-3}$, $x_{4m-2}$, $x_{4m-1}$, $x_{4m}$, the following three $d_2^2$'s may be considered:

(4.2)
$$d_{2(12)(34)}^2 = (x_{4m-3}-x_{4m-1})^2 + (x_{4m-2}-x_{4m})^2$$
$$d_{2(12)(43)}^2 = (x_{4m-3}-x_{4m})^2 + (x_{4m-1}-x_{4m-2})^2$$
$$d_{2(13)(24)}^2 = (x_{4m-3}-x_{4m-2})^2 + (x_{4m-1}-x_{4m})^2 \quad , (m = 1, 2, \ldots, n).$$

Of course three kinds of $D_2^2$ that are defined from the random variables ($X_{4m-3}$, $X_{4m-2}$, $X_{4m\_1}$, $X_{4m}$), (m=1, 2, … , n) in accordance with (4.2) are identically distributed according to (2.6). Each of the correlation coefficient between $D_{2(12)(34)}^2$, $D_{2(12)(43)}^2$ and $D_{2(13)(24)}^2$ is 2/7.

By the same discussion as in section 3, three kinds of $D_2^2$-test statistics $Q_{2(12)(34)}^2$, $Q_{2(12)(43)}^2$, $Q_{2(13)(24)}^2$ may be defined by

(4.3)
$$\begin{cases} Q_{2(12)(34)}^2 = \sum_{i=1}^{10} \frac{(f_{i(12)(34)}-F_i)^2}{F_i} \\ Q_{2(12)(43)}^2 = \sum_{i=1}^{10} \frac{(f_{i(12)(43)}-F_i)^2}{F_i} \\ Q_{2(13)(24)}^2 = \sum_{i=1}^{10} \frac{(f_{i(13)(24)}-F_i)^2}{F_i} \end{cases}$$

where $f_{i(12)(34)}$ is the observed number of components of $d_{2(12)(34)}^2$ falling into the i-th subinterval, $f_{i(12)(43)}$ and $f_{i(13)(24)}$ are analogous to $f_{i(12)(34)}$, and $F_i = n[F_2(a_i^2)-F_2(a_{i-1}^2)]$.

Using the random numbers (a), we calculated the correlation coefficients between $Q_{2(12)(34)}^2$, $Q_{2(12)(43)}^2$, $Q_{2(13)(24)}^2$ and $Q_3^2$.

The results are tabulated in Table 4.

|   | $Q_{2(12)(34)}^2$ | $Q_{2(12)(43)}^2$ | $Q_{2(13)(24)}^2$ | $Q_3^2$ |
|---|---|---|---|---|
| $Q_{2(12)(34)}^2$ | \ | -0.0002 | -0.0667 | 0.0198 |
| $Q_{2(12)(43)}^2$ |  | \ | -0.0326 | 0.1045 |
| $Q_{2(13)(24)}^2$ |  |  | \ | 0.0403 |
| $Q_3^2$ |  |  |  | \ |

Table 4: The correlation coefficients between $Q_3^2$ and three kinds of $Q_2^2$

From the results that they are at least nearly not correlated to each other, we may say that $D_3^2$-tests are different from $D_2^2$-tests, and that there are three different tests in $D_2^2$-tests.

## 5. Discussion

In Section 4, it was pointed out that $D_3^2$-tests were different from the other usual tests, and that there were three different tests in $D_2^2$-tests.

We can do the same discussion to $D_3^2$-tests. Suppose that $X_{6m-5}, X_{6m-4}, X_{6m-3}, X_{6m-2}, X_{6m-1}, X_{6m}$, are six consecutive random variables, $D_3^2$ is defined by, in accordance with (3.1),

(5.1) $\qquad D_3^2 = (X_{6m-5}-X_{6m-2})^2 + (X_{6m-4}-X_{6m-1})^2 + (X_{6m-3}-X_{6m})^2, \quad (m = 1, 2, \ldots, n).$

When we create two three-dimensional random vectors from six consecutive random variables and calculate the square of the Euclid distance between them we have fifteen ($_5C_2 \cdot _4C_2/3!$) different representations.

Let $\{D_3^2*\}$ be a set of them excluding $D_3^2$. A statistic in $\{D_3^2*\}$ has either two different terms from $D_3^2$ or all different terms from $D_3^2$. Let $\{D_3^2\,I\}$ be the former set that has six elements and $\{D_3^2\,II\}$ the later set that has eight elements. They are listed in the following form,

(5.2)
$\{D_3^2\,I\} = \{(123)(465), (123)(546), (123)(654), (124)(356),$
$\qquad (125)(436), (134)(265)\}$
$\{D_3^2\,I\} = \{(123)(564), (123)(645), (124)(365), (124)(536),$
$\qquad (124)(635), (125)(346), (134)(256), (125)(246)\},$

where (abc)(def) denotes $(X_{6(m-1)+a}-X_{6(m-1)+d})^2+(X_{6(m-1)+b}-X_{6(m-1)+e})^2+(X_{6(m-1)+c}-X_{6(m-1)+f})^2$ (a, b, c, d, e, f are six different integers, $1 \leq a, b, c, d, e, f \leq 6$).

Let $D_3^2\,I$ be an arbitrary element of $\{D_3^2\,I\}$ and $D_3^2\,II$ an arbitrary element of $\{D_3^2\,II\}$. As the terms in each of $D_3^2$, $D_3^2\,I$ and $D_3^2\,II$ are mutually independent, they have the same distribution function of (2.7) and have the means, variances and covariances given, respectively, by

(5.3)
$\qquad E(D_3^2) = E(D_3^2\,I) = E(D_3^2\,II) = 1/2$
$\qquad V(D_3^2) = V(D_3^2\,I) = V(D_3^2\,II) = 7/60$
$\qquad \text{Cov}(D_3^2, D_3^2\,I) = \text{Cov}(D_3^2\,I, D_3^2\,II) = 11/180$
$\qquad \text{Cov}(D_3^2, D_3^2\,II) = 1/30.$

And so, the correlation coefficients between these statistics are

(5.4)
$\qquad r(D_3^2, D_3^2\,I) = r(D_3^2\,I, D_3^2\,II) = 11/21$
$\qquad r(D_3^2, D_3^2\,II) = 2/7.$

Similar argument as in $D_2^2$-tests leads to two $D_3^2$-test statistics $Q_3^2$ and $Q_3^2$ II that are given in terms of $D_3^2$ and $D_3^2$ II , respectively.

Furthermore as mentioned above, since the mutual independence of $X_1, X_2, \ldots, X_{6n}$ implies that of $(X_1, X_2, X_3), \ldots, (X_{6n-2}, X_{6n-1}, X_{6n})$, we may consider that $D_3^2$-tests are also a test of three-dimensional random numbers.

At present, we have three-dimensional frequency tests for the test of three-dimensional random numbers in the unit cube, but they must have very large divided cores. If we divide the interval [0, 1] into p subintervals, the number of cores must be $p^3$. Then the accuracy of chi-square approximation may be poor, unless the size of random numbers is much larger than $p^3$. Additionally, they are the tests for the equiprobability only.

On the other hand, $D_3^2$-tests can be used for a comparatively small size of random numbers and may be considered for the test of independence.

## References

[1] F. Pesarin, L. Salmaso (2010). Permutation Tests for Complex Data, John Wiley & Sons, 158

[2] D. J. Bennett (1999). Randomness, Harvard University Press, 170-171

[3] L. Pardo, N. Balakrishnan, M. A. Gil (2011). Modern Mathematical Tools and Techniques in Capturing Complexity, Springer, 159-161.

[4] O. Miyatake, K. Wakimoto (1978). Random numbers and Monte Carlo methods, Morikita Company LTD